\documentclass[aps,prd,amsmath,amssymb,preprintnumbers,onecolumn,11pt,nofootinbib]{revtex4}

\usepackage[a4paper,left=20mm,right=20mm,top=25mm,bottom=25mm]{geometry}
\usepackage{mathrsfs}
\usepackage{graphicx,epsfig}
\usepackage{color}
\usepackage{fancyhdr}
\usepackage{multirow}
\usepackage{tikz}
\usepackage{feynmp}
\usepackage{mathpazo}
\usepackage{bbold}
\usepackage{natbib}
\usepackage{diagbox} 

\usepackage{amsfonts}
\usepackage{bm}
\usepackage{xcolor}

\definecolor{CP3}{cmyk}{0,0.88,0.77,0.40}

\newcommand\be{\begin{equation}}
\newcommand\ee{\end{equation}}
\newcommand\ba{\begin{eqnarray}}
\newcommand\ea{\end{eqnarray}}

\newcommand\wt{\widetilde}
\newcommand\e{{\rm e}}
\newcommand\de{\delta}

\newcommand\al{\alpha}
\newcommand\bt{\beta}
\newcommand\pa{\partial}
\newcommand\Om{\Omega}

\renewcommand{\(}{\left(}
\renewcommand{\)}{\right)}
\renewcommand{\[}{\left[}
\renewcommand{\]}{\right]}

\newcommand{\lm}{\lambda}
\newcommand{\sig}{\sigma}

\newcommand{\dm}{{m}}

\newcommand\Cp{C_{,\phi}}
\newcommand\Dp{D_{,\phi}}

\begin{document}

\title{\Large \color{CP3} Spherical collapse and cluster number counts in dark energy models disformally coupled to dark matter}
\author{Stharporn Sapa$^{1}$}
\email{stharporns57@email.nu.ac.th} 
\affiliation{\footnotesize $^{1}${The Institute for Fundamental Study \lq\lq The Tah Poe Academia Institute\rq\rq, \\Naresuan University, Phitsanulok 65000, Thailand}}
\author{Khamphee Karwan$^{1,2}$}
\email{khampheek@nu.ac.th}
\affiliation{\footnotesize $^{1}${The Institute for Fundamental Study \lq\lq The Tah Poe Academia Institute\rq\rq, \\Naresuan University, Phitsanulok 65000, Thailand}}
\affiliation{\footnotesize $^{2}${Thailand Center of Excellence in Physics, Ministry of Education,
Bangkok 10400, Thailand}}
\author{David F. Mota$^{3}$}
\email{d.f.mota@astro.uio.no}
\affiliation{\footnotesize $^{3}${Institute of Theoretical Astrophysics, University of Oslo, 
P.O. Box 1029 Blindern, N-0315 Oslo, Norway}}

\begin{abstract}
We investigate the effects of a disformal coupling between dark energy and dark matter in the predictions of the spherical collapse and its signatures in galaxy cluster number counts. We find that the disformal coupling has no significant effects on spherical collapse at high redshifts, and in particular during matter domination epoch. However, at lower redshifts, the extrapolated linear density contrast at collapse close to redshift $z \lesssim 1$ and overdensity at virialization can be strongly suppressed by a disformal coupling between dark energy and dark matter.
We also find that disformal coupling can have different imprints on cluster number counts compared with conformal coupling, such that the disformal coupling can strongly suppress the predicted number of clusters per redshift interval at $z > 0.1$
while enhance the number of cluster at $z < 0.05$.
Using the specifications of eROSITA survey, we find that the disformal coupling between dark energy and dark matter can be tightly constrained by cluster number counts.
\\[4mm]
{\footnotesize Keywords: couple dark energy, disformally coupled, spherical collapse, cluster number count}
\end{abstract}

\maketitle

\section{Introduction}

The observational data of Supernova Type Ia (SN Ia) \cite{Perlmutter-1, Perlmutter-2, Riess}, cosmic microwave background radiation \cite{CMB-1, CMB-2, CMB-3}, large-scale structure surveys (LSS) \cite{LSS-1, LSS-2} indicate that currently the universe is in the phase of accelerating expansion. In order to explain this phenomenon, we can assume that dynamics of the present universe is dominated by dark energy. The observations strongly indicate that the dark energy component corresponds to $70\%$ of the total energy density of the universe. Currently, properties of the dark energy are still unknown. There are many candidates to the dark energy, for instance cosmological constant \cite{Carroll et al. 1992} and scalar fields or quintessence models \cite{Ratra Peebles}.

Since the nature of dark matter and scalar field candidates to the dark energy are obscure,
we can explore properties of dark energy by supposing that dark energy can be coupled with dark matter \cite{amendola, DE1, euclid,DE2}. Although the couplings between dark energy and dark matter are well motivated from theories of particle physics and high energy physics \cite{coup1, coup2, coup4, coup7}, it could be more convenient  to gain our understanding in general how coupling between dark energy and dark matter can influence the observed universe using phenomenological forms of the coupling which have been variously proposed in literature \cite{coup3, coup6, coup8, coup9, coup10, coup11, coup12, coup14, coup15, coup16, coup17}, However, it is also interesting to study the coupling which arises from theoretical motivation. The couplings between dark energy and dark matter can be a consequence of frame transformation of the gravity action. The general transformation of frame which preserves causal structure of the theories is disformal transformation \cite{disf1, disf2, disf3, disf5}, and the coupling resulting from this transformation is known as disformally coupled. Influences of the disformal coupling between dark energy and dark matter on evolution of the background universe have been investigated in \cite{disf4, disf6, disfb1, disfb2}, while influences of the couplings on perturbations in the universe, i.e., CMB anisotropies, growth of structures, have been studied in \cite{disf7, disfp1, disfp2, deBruck:1709}.

In order to investigate structure formation of matter (ordinary and dark matter), Several authors have used N-body simulations to simulate formation of structure \cite{nbody1, nbody2, nbody3}. Other easier method for studying the structure formation and the influences of the dark energy to the overdense regions is the spherical collapse model \cite{spc4,spc5,spc6,spc0,spc1, spc2, spc3}. Using spherical collapse model together with Press-Schechter or Sheth-Torman formalism, cluster number counts of halos can be estimated. The cluster number counts can be used to study influence of the dark energy to overdense regions, test and discriminate among the dark energy models. In \cite{Wintergerst:2010}, evolution equations for non-linear perturbations for coupled dark energy models have been derived, and linear as well as non-linear density contrast at virialization for $\Lambda$CDM and the coupled dark energy models have been analyzed using spherical collapse. The influences of the coupling between dark energy and dark matter on cluster number counts are investigated in \cite{count1, count2}. In \cite{count3}, the effect of power law f(T) gravity to spherical collapse and cluster number counts is studied by comparing the results with $\Lambda$CDM model. The number density of galaxy clusters is investigated for various form of the potential of dark energy in \cite{count4}.

In this work, we investigate influences of disformal coupling between dark energy and dark matter on spherical collapse and cluster number counts. The covariant form of the disformal coupling between dark energy and dark matter is presented in section (\ref{sec2}). In section (\ref{sec3}), the necessary evolution equations for the background universe, linear and non-linear perturbations for studying spherical collapse in disformally coupled dark energy model are shown. Effects of disformal coupling on spherical collapse and cluster number counts are investigated in sections (\ref{sec4}) and (\ref{sec5}) respectively.
The conclusions are given in section (\ref{sec6}).

\section{Disformal coupling between dark energy and dark matter}
\label{sec2}
Under the disformal transformation the metric tensor is transformed as
\begin{equation}
\label{gbar-gen}
\bar{g}_{\mu\nu} = C(\phi)g_{\mu\nu} + D(\phi) \phi_{,\mu}\phi_{,\nu}\,,
\end{equation}
where subscript ${}_{,\mu}$ denotes partial derivative with respect to $x^\mu$. Here, we consider the case where the conformal and disformal coefficients $C$ and $D$ depend on $\phi$ only. The inverse of the above metric is
\ba
\bar{g}^{\mu\nu} 
= \frac{1}{C^2} ( C g^{\mu\nu} -D \phi^{,\mu}\phi^{,\nu})
\,.
\ea
Let us now suppose that the field $\phi$ plays a role of dark energy, so that the interaction between dark energy and dark matter can occur when the Lagrangian of dark matter depends on metric $\bar{g}_{\mu\nu}$ given in Eq.~(\ref{gbar-gen}). Thus we write the action for gravity in terms of metric $g_{\mu\nu}$ and write the action for dark matter in terms of $\bar{g}_{\mu\nu}$ as
\begin{equation}
S = \int d^4x \left\{ \sqrt{-g}\[ \frac{1}{2}R + P(\phi,X) \] 
+ \sqrt{-\bar{g}}\mathcal{L}_\dm(\bar{g}_{\mu\nu}, \psi, \psi_{,\mu})\right\}
\,,\
\label{action}
\end{equation}
where we have set reduced Planck mass $m_p = 1/\sqrt{8\pi G} = 1$, $P(\phi,X) \equiv X - V(\phi)$ is the Lagrangian of the scalar field, $X \equiv - \phi_{,\mu}\phi^{,\mu} / 2$, $V(\phi)$ is the potential of the scalar field and $\mathcal{L}_\dm$ is the Lagrangian of dark matter. We will neglect baryon and radiation in our consideration, because we will concentrate on the evolution of the universe during matter and dark energy dominated epochs and baryon has no direct coupling with dark energy. Varying this action with respect to $g_{\al\bt}$, we get
\begin{equation}
G^{\al\bt} = T^{\al\bt}_\phi + T^{\al\bt}_\dm\,,
\label{eom-g1}
\end{equation}
where $G^{\al\bt}$ is the Einstein tensor computed from $g_{\mu\nu}$, and the energy momentum tensor for scalar field and dark matter are defined in unbarred frame as
\ba
T^{\mu\nu}_{\phi} &\equiv& \frac{2}{\sqrt{-g}}\frac{\delta\(\sqrt{-g}P(\phi,X)\)}{\delta g_{\mu\nu}}\,, 
\label{tmn-p}\\
T^{\mu\nu}_{\dm} &\equiv& \frac{2}{\sqrt{-g}}\frac{\delta\(\sqrt{-\bar{g}}\mathcal{L}_{\dm}\)}{\delta g_{\mu\nu}}\,.
\label{tmn-m}
\ea
Using these definitions of the energy momentum tensor, Eq.~(\ref{eom-g1}) implies
$\nabla_\al (T^{\al\bt}_\phi + T^{\al\bt}_\dm) =0$. However, we see that the energy momentum tensors of dark energy and dark matter do not separately conserve because the Lagrangian of dark matter depends on field $\phi$. The energy momentum tensor in the barred frame is related to that in the unbarred frame defined in Eq.~(\ref{tmn-m}) as
\be
T^{\al\bt}_\dm 
= \frac{\sqrt{-\bar{g}}}{\sqrt{-g}}\,
 \frac{\de \bar{g}_{\rho\sig}}{\de g_{\al\bt}}\, \frac{2}{\sqrt{-\bar{g}}} \frac{\delta\(\sqrt{-\bar{g}}\mathcal{L}_{\dm}\)}{\delta \bar{g}_{\rho\sig}} 
\quad
= \frac{\sqrt{-\bar{g}}}{\sqrt{-g}}\frac{\de \bar{g}_{\rho\sig}}{\de g_{\al\bt}}\bar{T}^{\rho\sig}_\dm\,.
\label{bar-unbar}
\ee
Varying the action (\ref{action}) with respect to $\phi$, we obtain
\ba
\phi^{,\al}_{;\al} - V_{,\phi} 
&=& 
\nabla_\bt\(\frac{\sqrt{-\bar{g}}}{\sqrt{-g}} \bar{T}^{\al\bt}_\dm D \phi_{,\al}\)
- \frac 12 \frac{\sqrt{-\bar{g}}}{\sqrt{-g}} \bar{T}^{\al\bt}_\dm\(\Cp g_{\al\bt} + \Dp\phi_{,\al}\phi_{,\bt}\)
\nonumber\\
&\equiv& Q\,,
\label{gkg-q1}
\ea
where ${}_;$ denotes the covariant derivative and a subscript ${}_{,\phi}$ denotes derivative with respect to $\phi$. Multiplying the above equation by $\phi^{,\lm}$, we get
\be
Q\phi^{,\lm} 
=
\nabla_\al T^{\al\lm}_\phi = - \nabla_\al T^{\al\lm}_\dm\,.
\label{dt-q}
\ee
According to \cite{disfp1,disfb2}, the barred quantities in the interaction term $Q$ can be written in terms of unbarred quantities, so that $Q$ can be written as
\ba
Q
&=& 
\frac 1{2C^2}\[\(C D_{,\phi} - 2 C_{,\phi} D\) \phi_{,\al}\phi_{,\bt} T^{\al\bt}_\dm 
\right.
\nonumber\\
&&  \left.
- C\( C_{,\phi} g_{\al\bt} T^{\al\bt}_\dm 
- 2 D ( \phi_{;\al\bt} T^{\al\bt}_\dm\ + \phi_{,\al} \nabla_\bt T^{\al\bt}_\dm\ ) \)\]\,.
\label{q-ub}
\ea

\section{Evolution equations}
\label{sec3}
In this section, we will present necessary evolution equations for studying spherical collapse in coupled dark energy model.
The spherical collapse model is a simple tool to  follow the growth of non-linear overdensity of matter inside spherical region embedded in the background universe where in the usual case the overdensity inside the region is assumed to be uniformly distributed.
The uniform distribution of overdensity can be described by the top hat density profile  in which the overdensity $\rho(t, r') = \rho(t)$ when the distance from the center of the region $r'$ does not exceed radius of the region, but zero otherwise.
Evolution of the radius of the region is governed by magnitude of the overdensity inside the region $\rho(t)$ which is time-dependent.
In the spherical collapse model, dynamics of the spherical region containing overdensity $\rho$ obey the following ``'Friedmann equation'':
\ba
\label{eq:sc_f1b} 
\(\frac{\dot r}{r}\)^2 &=& \frac 13 \rho - \frac{K}{r^2} = \frac{1}{3} \sum_{\alpha} \rho_{\alpha} - \frac{K}{r^2}\,,
\\
 \label{eq:sc_f2b} 
\frac{\ddot r}{r} &=&  - \frac 16 \(\rho + 3 p\)
= - \frac{1}{6} \sum_{\alpha} \(\rho_{\alpha} + 3 p_{\alpha}\)\,,
\ea
where a dot denotes derivative with respect to time $t$,
index $\alpha$ runs over the matter component evolving the evolution of the region,
$r$ is the radius of the spherical region,
$p$ is the total pressure of the matter inside the region,
$\rho_\alpha$ and $p_\alpha$ are the energy density and pressure of the $\alpha$th  component of matter.
The parameter $K$ defines the critical energy density which is the minimum density required to make the region collapse.
Practically, processes of spherical collapse can be studied by computing the evolution of the overdensity inside the spherical region from evolution equation for non-linear density perturbation.
For the case where the dark energy is in the form of scalar field with canonical kinetic  term and has no direct interaction with dark matter,
the dark energy is not expected to cluster on small scales because its effective sound speed for density perturbations equals to speed of light.
In coupled dark energy models, the interaction between dark energy and dark matter may alter the effective sound speed of density perturbation of dark energy,
and consequently the dark energy can cluster on small scales.
However, According to Eq.~(\ref{q-ub}),
the effective sound speed of dark energy is not modified by the disformal coupling because the coupled term does not contain the term that is proportional to second order spatial derivative of $\phi$ when dark matter is a pressureless perfect fluid.
Hence, only the dark matter and baryon density contribute to the dynamics of the spherical collapse in this case.
Since the interaction between dark energy and dark matter modifies the growth rate of the density perturbation of dark matter compared with baryon which is not expected to have direct interaction with dark energy,
the overdensity of dark matter and baryon collapse with different rate in coupled dark energy models.
As a result, the ratio of baryon density to dark matter density in the clusters is influenced by the interaction between dark energy and dark matter \cite{Mainini:2006zj, Le Delliou:12}.
However, to avoid the complexities of the collapsing processes,
we do not consider this influence and ignore baryon in our consideration similar to the analysis in \cite{Wintergerst:2010, coup15}.
Before introducing the evolution equation for non-linear perturbation in disformally coupled dark energy model,
let us first present evolution equation for the background universe in the following section.

\subsection{Evolution equations in the FLRW universe}
\label{flrw}
Using the Friedmann-Lema\^{i}tre-Robertson-Walker (FLRW) line element, 
\be
ds^2 = - dt^2 + a^2(t)\de_{ij}dx^i dx^j\,,
\label{ds-flrw}
\ee
Eq.~(\ref{eom-g1}) yields
\be
H^2 \equiv \(\frac{\dot a}{a}\)^2 = \frac 13 \[\rho_\dm + \frac 12 (\dot\phi)^2 + V(\phi)\]\,,
\label{h2}
\ee
where $\rho_\dm$ is the energy density of dark matter. Furthermore, the interaction terms $Q$ in Eq.~(\ref{q-ub}) becomes
\ba
Q
&=& 
\frac 1{2C^2} \Big[ 2 \( C D_{,\phi} - 2 C_{,\phi} D \) X \rho_\dm \nonumber\\
&& + C \Big( C_{,\phi} \rho_\dm + 2 D \big( \ddot\phi \rho_\dm + \dot\phi \( \dot\rho_\dm + 3 H \rho_\dm \) \big) \Big) \Big]\,,
\label{q0-start}
\ea
In the above expression, all quantities are evaluated in the background universe, such that $X \equiv (\dot\phi)^2 / 2$. Inserting this expression for the interaction terms into Eq.~(\ref{dt-q}), we can compute the expression for $\dot\rho_\dm$. Substituting the expression for $\dot\rho_\dm$ back into Eq.~(\ref{q0-start}), and then inserting the result into Eq.~(\ref{gkg-q1}), we get \cite{disfp1,disfb2}
\ba
 \ddot\phi + 3 H \dot\phi + V_{,\phi} 
&=& \frac{4 C C_{,\phi} D X + C^2 \big[ 2 D (3 H \dot{\phi} + V_{,\phi}) - (C_{,\phi} + D_{,\phi} X )\big]}
{2 C^2 [C + D (\rho_\dm - 2 X)]} \rho_m \nonumber\\
&\equiv& - Q_0\,,
\label{defq0}
\ea
Using the above results, it can be shown that the evolution equations for $\rho_m$ is
\be
\dot\rho_m + 3 H \rho_m = Q_0 \dot\phi\,.
\label{dotrhom}
\ee

\subsection{Evolution equations for the perturbations on small scales}
\label{smallpert}
In order to obtain evolution equations for spherical collapse, we firstly compute the evolution equations for density contrast $\de_m \equiv \de\rho_m / \rho_m$ and velocity perturbation $v_m^i$ for dark matter on small scales. 
Since the dark matter is usually modeled by pressureless perfect fluid which has no anisotropic perturbation,
and disformal coupling between dark energy and dark matter cannot generate anisotropic perturbations \cite{disfp1},
the line element can be written in the weak field  limit as
\be
ds^2 = - (1 + 2\Phi) dt^2 + a^2(1 - 2 \Phi)\de_{ij}dx^i dx^j\,.
\label{ds-pert}
\ee
On small scales, the Einstein theory of gravity converges to Newtonian limit in which the component $\mu\nu = 00$ of the perturbed Einstein equation yields \cite{amendola_2004, pettorino_baccigalupi_2008, Wintergerst:2010}
\be
\pa_i\pa^i \Phi \equiv \nabla^2\Phi \simeq \frac 12 \delta\rho_m\,.
\label{p2phi}
\ee
On sufficiently small scales, we have $\nabla^2 \de\phi \gg \de\ddot{\phi}, H\de\dot{\phi}$, where $\de\phi$ is the perturbations in $\phi$, so that Eq.~(\ref{gkg-q1}) yields \cite{amendola_2004, pettorino_baccigalupi_2008, Wintergerst:2010, disfp2}
\be
\nabla^2 \delta\phi = \de Q\,,
\label{p2dp}
\ee
where $\de Q$ is the perturbation in the coupling term $Q$.
Since  $\de Q$ is time-dependent,
the above equation suggests that the perturbations in dark energy field as well as density perturbation of dark energy are also time-dependent.
However, this temporal variation of the field perturbations is negligible compared with the spatial variation on small scales on which spherical collapse model is operated.
In the situation where the dark energy does not cluster on small scales,
the density perturbations of dark energy are always negligible compared with density perturbations of dark matter on small scales inside the collapsing region.
Hence, the temporal variation of perturbations in dark energy field  does not affect the metric perturbation given in Eq.~(\ref{p2phi})
and the perturbations in coupling term $\de Q$.
As a result,  the non-linear growth of matter perturbation inside the collapsing region weakly depends on temporal variation of perturbations in dark energy field.
The dark energy mainly influences processes of spherical collapse through the background evolution in our consideration.
The evolution equations for $\de_m$ and $v_m^i$ can be computed from Eq.~(\ref{dt-q}) by using the energy momentum tensor for dark matter of the form
\be
T_m{}^{\al\beta} \equiv \(\rho_m + \de\rho_m\) U^\al U^\beta\,,
\label{tmn-pf}
\ee
where $\rho_m$ is the background energy density of dark matter, $U^\al = (1 - \Phi, v_m^i)$ is the four velocity of dark matter and $v_m^i$ is the 3D comoving velocity of dark matter. Applying the small scales approximation to Eq.~(\ref{dt-q}), the component $\lm = 0$ of Eq.~(\ref{dt-q}) on small scales becomes
\ba
\dot\de_m  =
- (1 + \de_m)\pa_i v_m^i
- v_m^i \pa_i \de_m
- \wt{Q_0} \dot\phi \de_m
+ \frac{\de Q}{\rho_m}\dot\phi\,,
\label{dec-nl}
\ea
where $\de_m \equiv \de\rho_m / \rho_m$ and
\be
\wt{Q_0} \equiv \frac{Q_0}{\rho_m}
= - \frac{4 C_{,\phi} D X -  C \bigl(C_{,\phi} - 2 D (3 \dot{\phi} H + V_{,\phi}) + 2 D_{,\phi} X\bigr)}
{2 C \bigl[C + D (\rho_\dm - 2 X)\big]}\,.
\label{q0-cdp}
\ee
Similarly, the component $\lm = i$ of Eq.~(\ref{dt-q}) on small scales becomes
\begin{equation}
\dot v_m^i=
- \(2 H + \wt{Q_0} \dot\phi\) v_m^i
-  v_m^j \pa_j v_m^i
- \(\pa^i\Phi + \wt{Q_0} \pa^i\de\phi\)\,.
\label{dotvc}
\end{equation}
The perturbations in the interaction term $\de Q$ appearing in Eqs.~(\ref{p2dp}) and (\ref{dec-nl}) can be computed by applying the small scales approximation to Eq.~(\ref{q-ub}). On small scales, the dominant contributions in $\de Q$ computed from Eq.~(\ref{q-ub}) are
\be
\frac{\de Q}{\rho_m} =
\frac{D \dot{\phi}}{C}\[
\dot\de_m + (1 + \de_m)\pa_i v_m^i + v_m^i \pa_i \de_m
\]
+ \wt{Q_0} \de_m\,.
\label{deqraw1}
 \ee
Substituting the above equation into Eq.~(\ref{dec-nl}), we get
\be
\dot\de_m  = - (1 + \de_m)\pa_i v_m^i - v_m^i \pa_i \de_m\,. 
\label{dem-1}
\ee
Inserting Eq.~(\ref{dem-1}) into Eq.~(\ref{deqraw1}), the term $\dot{\de}_m$ can be eliminated and the resulting interaction term is
\be
\frac{\de Q}{\rho_m} =
\wt{Q_0} \de_m\,.
\label{deqfin1}
 \ee
Hence, Eq.~(\ref{p2dp}) becomes
\be
\nabla^2 \delta\phi = \wt{Q_0} \de\rho_m\,.
\label{p2dp-1}
\ee
In order to derive the non-linear evolution equation for $\de_m$, we apply the assumption for top hat density profile \cite{Wintergerst:2010} to Eqs.~(\ref{dem-1}) and (\ref{dotvc}), and differentiate these equations with respect to time. After eliminating $\pa_i \dot v_m^i$ from the resulting equations, we get
\ba
\ddot\de_m
&=& 
- \(2 H + \wt{Q_0}\dot\phi\) \dot\de_m
+ \frac 43 \frac{\dot\de_m^2}{1 + \de_m}
\nonumber\\
&&
+ \(1 + \de_m\)
\(\nabla^2\Phi + \wt{Q_0} \nabla^2\de\phi\)\,.
\label{ddde1}
\ea
Substituting Eqs.~(\ref{p2phi}) and (\ref{p2dp}) into the above equation,
one can see that the evolution of $\de_m$ on small scales is independent of length scales which is compatible with top hat density profile.
In spherical collapse model, the non-linear growth of $\de_m$ on small scale inside the collapsing region can depend on the length scales when sound speed of dark energy is much smaller than unity but significantly larger than zero \cite{Basse:10}.
For this case, the overdensity inside the collapsing region is not compatible with the top hat density profile, and the calculation in the standard spherical collapse model is required to be modified.

To connect the evolution of the radius $r$ of the top hat region containing non-linear density contrast $\delta_{m}$ with the evolution of $\delta_m$,
we use the assumption that the energy-momentum transfer between dark energy and dark matter due to direct interaction modifies effective mass of dark matter particle rather than changes the number of dark matter particle.
Hence, the top hat number density  $n$ of dark matter inside the region with radius $r$ relates to the number density $\bar{n}$ of  dark matter in the background universe as $n \propto \bar{n} (a/r)^3$.
Let $M(t)$ be  an averaged effective mass of dark matter particle inside the top hat region,
and $\bar{M}(t)$ be an effective mass of dark matter particle in the background universe,
the ratio between the overdensity inside the top hat region and the energy density of dark matter in the background universe can be written as
$\rho_m / \bar\rho_m = M(t) n / (\bar{M}(t) \bar{\rho}_m) \propto (M(t)/ \bar{M}(t)) (a/r)^3$.
In the standard spherical collapse model,
the ratio $M(t) / \bar{M}(t)$ can be computed by integrating the conservation equations with coupling terms for $\rho_m$,
while the evolution of $r$ still obeys Eqs.~(\ref{eq:sc_f1b}) and (\ref{eq:sc_f2b}) \cite{Wintergerst:2010, Tarrant:12}.
Alternatively, the influence of the coupling between dark energy and dark matter on the evolution of the radius $r$ can be presented in terms of the extra force in the evolution equation for $r$ if the contribution from  dark energy interaction is not taken into account in the relation between overdensity and radius of the top hat region.
Setting $M(t)/\bar{M}(t) = 1$,
we have\cite{Wintergerst:2010}
\be
1 + \delta_m = (1 + \delta_{m,in})\,\left(\frac{a}{r}\right)^3\,, 
\label{masscons}
\ee
where $\delta_{m,in} \ll 1$ is the initial values of $\de_m$ and we have set $r = a$ initially.
Differentiating the above equation with respect to time
and comparing the result with Eq.~(\ref{ddde1}),
we will obtain the modified version of Eq.~(\ref{eq:sc_f2b}) which contains the extra force terms associated with the coupling between dark energy and dark matter.
The derivation of the evolution equation for $r$ with extra force term from Eqs.~(\ref{masscons}) and (\ref{ddde1})
is  performed in  \cite{Wintergerst:2010}.
This  evolution equation is equivalent to that is used in \cite{Mainini:2006zj}.
Comparison between the standard spherical model and  alternative approach is presented in \cite{Wintergerst:2010, Tarrant:12}.

\section{Spherical collapse}
\label{sec4}
\begin{figure}
\includegraphics[height=0.4\textwidth, width=0.9\textwidth,angle=0]{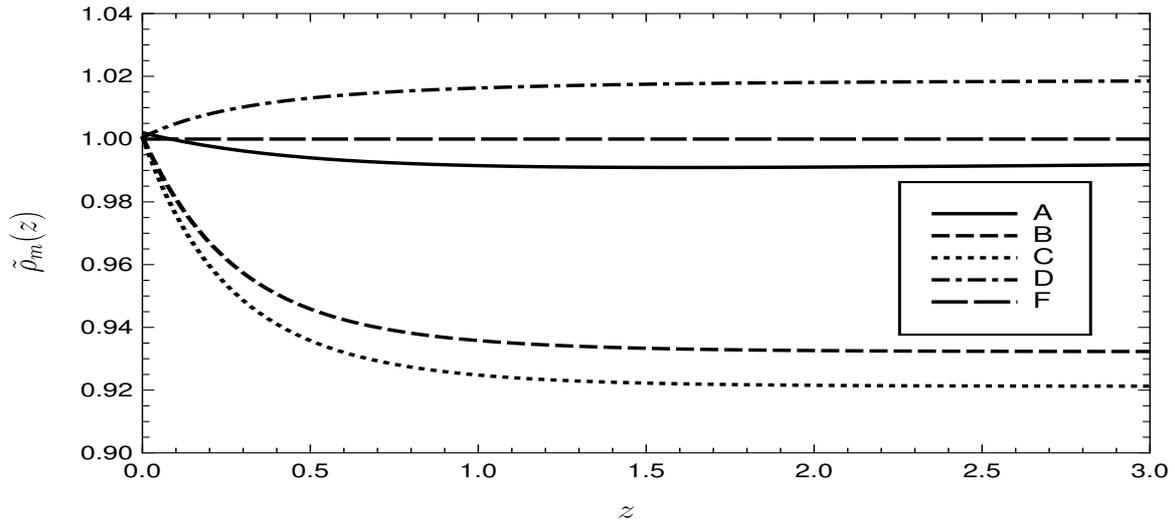}
\caption{\label{fig:1}
Plots of $\tilde\rho_m \equiv a^3\rho_m / \rho_{m0}$ as a function of redshift $z$. The lines A, B, C, D, and F represent the models A, B, C, D and F in Table \ref{modelparam} respectively.
}
\end{figure}

\begin{figure}
\includegraphics[height=0.4\textwidth, width=0.9\textwidth,angle=0]{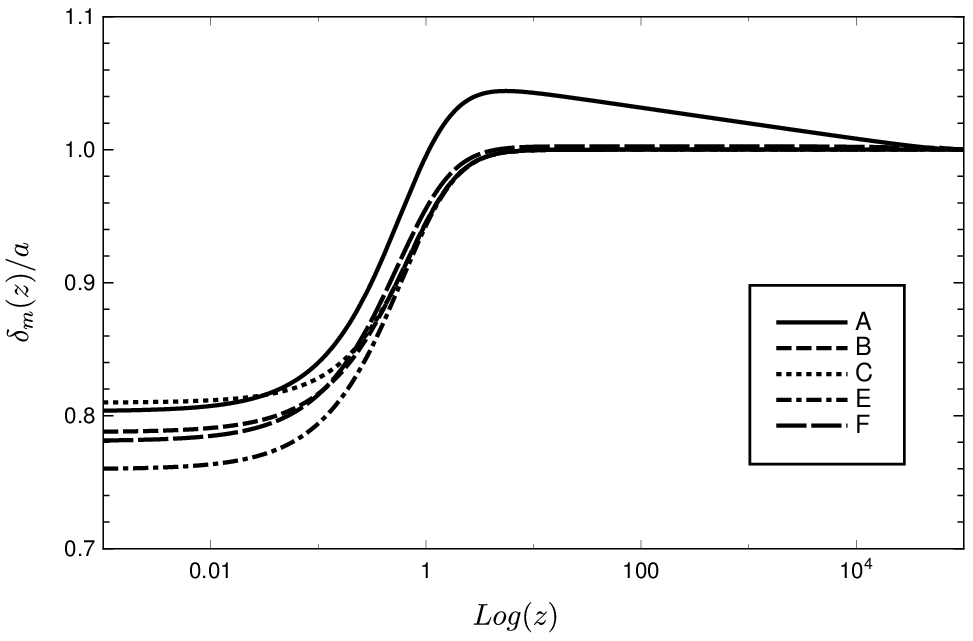}
\caption{\label{fig:2}
Plots of $\de_m / a$ as a function of redshift $z$.
The lines A, B, C, E and F correspond to models A, B, C, E and F in Table \ref{modelparam} respectively.
}
\end{figure}

In order to perform further study, we use
\be
C = \e^{\lm_1 \phi}\,,
\quad
D = M_d^{-4} \e^{\lm_2 \phi}\,,
\quad
V  = M_v^4 \e^{\lm_3\phi}\,,
\label{def-cd}
\ee
where $\lm_1$, $\lm_2$ and $\lm_3$ are the dimensionless constant parameters, while $M_d$ and $M_v$ are the constant parameters with dimension of mass. Using the dimensionless variables
\be
x_1^2 \equiv \frac{\dot\phi^2}{6 H^{2}}\,,
\quad 
x_2 \equiv \frac{V}{3 H^{2}}\,,
\quad
x_3 \equiv \frac{D H^{2}}{C}\,,
\label{def-dyna}
\ee
we can write the evolution equations for the background universe given in section \ref{flrw} in the autonomous form as \cite{disfb1,disfb2}
\ba
x_1' &=& 
\frac {1}{2} \Big(
x_1 \(3 x_1^2 - 3 x_2 +1\) - 2 (\sqrt{3/2} \lambda_3 x_2 + 2 x_1) 
\Big)
\nonumber\\
&&
- \frac{\sqrt{3}}{2\sqrt{2}}\(x_1^2 + x_2 - 1\)
\frac{\lambda_1 \big(12 x_1^2 x_3 - 1\big)
-6 x_3 \big(\lambda_2 x_1^2 - \sqrt{6} x_1 - \lambda_3 x_2 \big)
}
{1 - 3 x_3 \(3 x_1^2 + x_2 - 1\)}\,,
\label{x1p}\\
x_2'
&=&
x_2 \big(\sqrt{6} \lambda_3 x_1 + 3 x_1^2-3 x_2 + 3\big)\,,
\label{x2p}\\ 
x_3'
&=& 
- x_3 \big[
3 x_1^2  + \sqrt{6} (\lambda_1 - \lambda_2) x_1-3 x_2+3
\big]\,,
\label{x3p}
\ea
where a prime denotes derivative with respect to $N \equiv \ln a$. The evolution of the background universe is completely described by the above equations. The density parameter of dark matter $\Omega_m$ is related to the above dimensionless variables through Eq.~(\ref{h2}) as
\be
1= x_1^2 + x_2 + \Om_m\,.
\label{omm}
\ee
Using the definition in Eq.~(\ref{def-dyna}), $x_3$ can be expressed in terms of $x_2$ as
\begin{align}
x_3 &= \frac{D M_p^2 H^{2}}{C} = \frac{M_p^2 H^2}{M_d^4} \e^{(\lm_2- \lm_1)\phi / M_p} \nonumber\\
& = 
\frac{M_p^2 H_0^2 E^2}{M_d^4} \(\frac{3 M_p^2 H_0^2}{M_v^4} E^2 x_2\)^{(\lm_2 - \lm_1) / \lm_3}\,,
\label{x22x3}
\end{align}
where $E \equiv H / H_0$ and $H_0$ is the present value of the Hubble parameter. The reduced Planck mass is restored in the above expression to avoid confusion. From observations, we have $M_p^2 H_0^2 \simeq 2.7 \times 10^{-47}$ GeV${}^4$ $\simeq 27$ meV${}^4$. We choose $M_d = M_v \simeq 1 / 0.55$ meV \cite{disfp2}. It follows from Eqs.~(\ref{x1p}) -- (\ref{x3p}) and (\ref{x22x3}) that the evolution equations for $x_1$, $x_2$ and $E$ also form a complete set of evolution equations for the background universe. Since the evolution of $E$ is required to compute cluster number counts in the next section, we solve the evolution equations for $x_1$, $x_2$ and $E$ instead of those for $x_1$, $x_2$  and $x_3$. The evolution equation for $E$ can be computed by differentiating Eq.~(\ref{h2}) with respect to $N$ yielding the result
\be
\frac{E'}{E} = 
\frac{\dot{H}}{H^2} = 
\frac 32 \(x_2 - 1 - x_1^2\)\,.
\label{ep}
\ee
In order to numerically solve evolution equations for the background universe, we set initial conditions for $x_1$ and $x_2$ such that the density parameter of dark energy $\Om_d$ takes value 0.7 at present and equation of state parameter of dark energy $w_d$ lies within the range $-1 < w_d < -0.9$. The initial values for $E$ is chosen from the requirement that $E = 1$ at present. We now discuss evolution of the background universe. The coupling term $\wt{Q_0}$ can be written in terms of dimensionless variables as
\be
\wt{Q_0} = 
\frac{\lambda_1 - 6 \left(2 \lambda_1-\lambda_2\right) x_3 x_1^2 - 6 \lambda_3 x_2 x_3 - 6 \sqrt{6} x_3 x_1}
{6 \left(1 - 3 x_1^2 - x_2\right) x_3 + 2}\,.
\label{q1auto}
\ee
Inserting Eq.~(\ref{q1auto}) into Eq.~(\ref{dotrhom}), we get
\be
\dot\rho_m + 3 H \(1 - \sqrt{\frac 23} \wt{Q_0} x_1\)\rho_m = 0\,.
\label{dotrhom-mo}
\ee
During matter domination, we have $x_1, x_2 \ll 1$, so that Eq.~(\ref{q1auto}) becomes
\be
\wt{Q_0} = 
\frac{\lambda_1}
{2 + 6 x_3} \,.
\label{q1matter}
\ee
This suggests that during matter domination, the effects of the conformal coupling quantified by $\lm_1$ are suppressed by the amplitude of disformal coefficient quantified by $x_3$. For the case where $0 < \lm_1,|\lm_2|, |\lm_3| \lesssim 1$, and $M_d \sim 1$meV, the expressions in Eq.~(\ref{x22x3}) gives $x_3 \sim E^2 \gg 1$ during matter domination. Hence, the disformal coupling can strongly suppress effects of the conformal coupling as well as the total magnitude of the coupling during matter domination. In addition to the suppression due to the disformal coupling, the effect of the coupling term in Eq.~(\ref{dotrhom-mo}) can be reduced if dark energy slowly evolves, i.e., $x_1 \ll 1$. The magnitude of $x_1$  is mainly controlled by slope of the potential of dark energy which depends on the parameter $\lm_3$. The other main different feature of the disformally coupled models compared with pure conformally coupled models is that the disformal coupling can lead to large magnitude of the coupling between dark energy and dark matter at late time while the coupling is negligible during matter domination. It follows from the Eq.~(\ref{q1auto}) that for $\lm_3 < 0$, the third term in the numerator can enhance the magnitude of the coupling when $x_2 \sim 0.7$ and $x_3 \sim 1$ at late time. For the pure conformally coupled model, $\wt{Q_0} = \lm_1 /2$ during both dark energy and matter domination, so that the evolution of the universe during matter domination may become unphysical if $\lm_1 \sim 1$. During matter domination, if $\lm_1 \sim 1$ and $\rho_d \ll \rho_m$, where $\rho_d$ is the background energy density of dark energy, the last two terms on the LHS of Eq.~(\ref{defq0}) will be much smaller than the coupling term on the RHS. Consequently, the dark energy field $\phi$ will be strongly driven by ``external force'' $Q_0$, and therefore matter dominated epoch will stop quickly and usuall acceleration epoch cannot start properly. However, if $\rho_d$ is not too small compared with $\rho_m$ during matter domination, the universe can evolve properly although $\lm_1 > 1$.
This situation occurs, for example, when dark energy establishes scaling solution during matter domination in the quintessence model with exponential potential (see e.g.~\cite{count1}).
In this case, dark energy in the matter dominated epoch can give a significant contribution to the spherical collapse and cluster number counts \cite{doran}. In our consideration, we suppose that dark energy slowly evolves throughout the whole evolution of the universe, so that $\rho_d \ll \rho_m$ during matter domination.
\begin{table}
\begin{center}
\begin{tabular}{|c|c|c|c|c|c|c|}
\hline
\diagbox{$\lambda_i$}{Model}  & $A$ & $B$ & $C$ & $D$ & $E$ & $F$ \\
\hline
$\lambda_1$ & 0.1 & 0.1 & 0.1 & 0.1 & 0 & - \\
\hline
$\lambda_2$ & 0 & 0 & -2 & -1 & 0 & - \\
\hline
$\lambda_3$ & -1 & -1 & -1 & -0.1 & -1 & - \\
\hline
 & conformal & \multicolumn{3}{|c|}{disformal} & uncoupled & $\Lambda$CDM \\
\hline
\end{tabular}
\caption{\label{modelparam}
Values of parameters $\lambda_1, \lambda_2$ and $\lambda_3$ for each model.
Model A , E and F are pure conformally coupled, uncoupled and $\Lambda$CDM models respectively.
Model B, C, D are disformally coupled models.
For disformally coupled models, $M_d$ defined in Eq.~(\ref{def-cd}) is set to $M_d \simeq 1 / 0.55$ meV,
while  $M_d = 0$ for other models.
}
\end{center}
\end{table}
To check influences of the coupling term $\wt{Q_0}$ on the evolution of $\rho_m$, we plot in Figure~(\ref{fig:1}) evolution of $\tilde\rho_m \equiv a^3\rho_m / \rho_{m0}$. Here, $\rho_{m0}$ is the present value of $\rho_m$. From the plot, we see that $\rho_m \propto a^{-3}$ at high redshifts. For a fixed $\rho_{m0}$, $\rho_m$ at a given redshift decreases when $\lm_1$, $-\lm_2$ increase for $\lm_3 = -1$, because $\wt{Q_0}$ increases in this situation. It follows from Eq.~(\ref{q1auto}) that $\wt{Q_0}$ increases when $\lm_1$ increases. According to Eq.~(\ref{x22x3}), a negative $\lm_2$ can enhance $x_3$ at late time for negative $\lm_3$, because $3 M_p^2 H_0^2 E^2 x_2 / M_v^4 > 1$. In the case where $\lm_3 = -1$, the third term in the numerator of Eq.~(\ref{q1auto}) can give a dominant contribution when $-\lm_2$ increases due to an enhancement of $x_3$.
From Eq.~(\ref{x22x3}), we see that the increasing of $\lm_3$ from negative value towards zero can enhance $x_3$ at late time, consequently $\wt{Q_0}$ can become negative due to a large contribution from the fourth term in the numerator of Eq.~(\ref{q1auto}). When $\wt{Q_0}$ becomes negative, $\rho_m$ will decay faster than $a^{-3}$ as presented by line D in the Figure~(\ref{fig:1}).
To study how the disformal and conformal coupling influence growth of density perturbations, we insert Eqs.~(\ref{p2phi}) and (\ref{p2dp-1}) into Eq.~(\ref{ddde1}), and then write the resulting equation in terms of the dimensionless variables as
\ba
\de_m''
&=& 
- \(\frac 12 \(1 + 3x_2 - 3 x_1^2\) 
+ \sqrt{6} \wt{Q_0} x_1\) \de_m'
+ \frac 43 \frac{(\de_m')^2}{1 + \de_m}
\nonumber\\
&&
+ \frac 32 \(1 - x_1^2 - x_2\)\(1 + \de_m\)
\(1 + 2 \wt{Q_0}^2\)\de_m\,.
\label{depp1}
\ea
The linearized version of this equation is
\begin{align}
\de_m''
=& 
- \(\frac 12 \(1 + 3x_2 - 3 x_1^2\) 
+ \sqrt{6} \wt{Q_0} x_1\) \de_m' \nonumber\\
& + \frac 32\(1 - x_1^2 - x_2\)
\(1 + 2 \wt{Q_0}^2\)\de_m\,.
\label{depp1-lin}
\end{align}
During matter domination, $\wt{Q_0}$ obeys the approximation given in Eq.~(\ref{q1matter}). Hence, for the case where $x_3 = 0$, i.e., pure conformally coupled model, Eq.~(\ref{depp1-lin}) is satisfied by the following growing solution:
\be
\de_m \propto \e^{p N}\,,
\quad\mbox{where}\quad
p = - \frac 14 + \frac 14 \sqrt{25 + 12\lm_1^2}\,.
\label{conformal-gw}
\ee
This shows that the conformal coupling can enhance the growth of $\de_m$ during matter domination. As discuss above, this enhancement can be disappeared due to disformal coupling which is in agreement with the plots of $\de_m /a$ in Figure~(\ref{fig:2}). In the plot, the ratio $\de_m/a$ for pure conformally coupled model at a given redshift during matter domination is larger than that for uncoupled model, and this ratio for disformally coupled and uncoupled models are not significantly different.
The enhancement of the growth rate of $\de_m$ during matter domination due to the conformal coupling between dark energy and dark matter is clearly followed from Eq.~(\ref{conformal-gw}).
Nevertheless,  this enhancement is not visible in a plot of $\de_m$ versus a scale factor as presented in figure 7 in \cite{disfp1},
so that we plot $\de_m /a$ rather than $\de_m$ in figure (\ref{fig:2}).
At late time, the ratio $\de_m/a$ for disformally coupled model can decrease slower than that in the uncoupled and pure conformally coupled models which is in agreement with \cite{disfp2}. The decreasing rate of $\de_m/a$ for coupled models at late time depends on $\wt{Q_0}^2$ term which controls ``growing rate'' of $\de_m$ in Eq.~(\ref{depp1-lin}). In Figure~(\ref{fig:2}), we see that the dependence of decreasing rate of $\de_m/a$ on $\lm_1$, $\lm_2$ and $\lm_3$ can be understood from the dependence of $\wt{Q_0}$ on these parameters at late time discussed above.
In the spherical collapse model, a region of overdensity with radius $r$ expands at initial stage due to the expansion of the background universe because magnitute of overdensity is small .
However, the expansion of the radius $r$ is slower than the Huble expansion,
because the gravitational attraction of the overdensity inside the region.
As a result, the overdensity can grow non-linearly and therefore gravitational attraction will be strong enough to stop the expansion of the radius $r$.
This is a turn around stage at which the radius $r$ is maximum and starts to reduce  due to gravitational attraction.
In the process of structure formation, a region of overdensity will not collapse to a singularity at $r = 0$ and $\de_m = \infty$
due to a balance between kinetic and gravitational potential of the region.
This balance is a virialization of a collapsing region.
Roughly speaking, structures are formed when the virialization is taken place.

\begin{figure}
\includegraphics[height=0.4\textwidth, width=0.9\textwidth,angle=0]{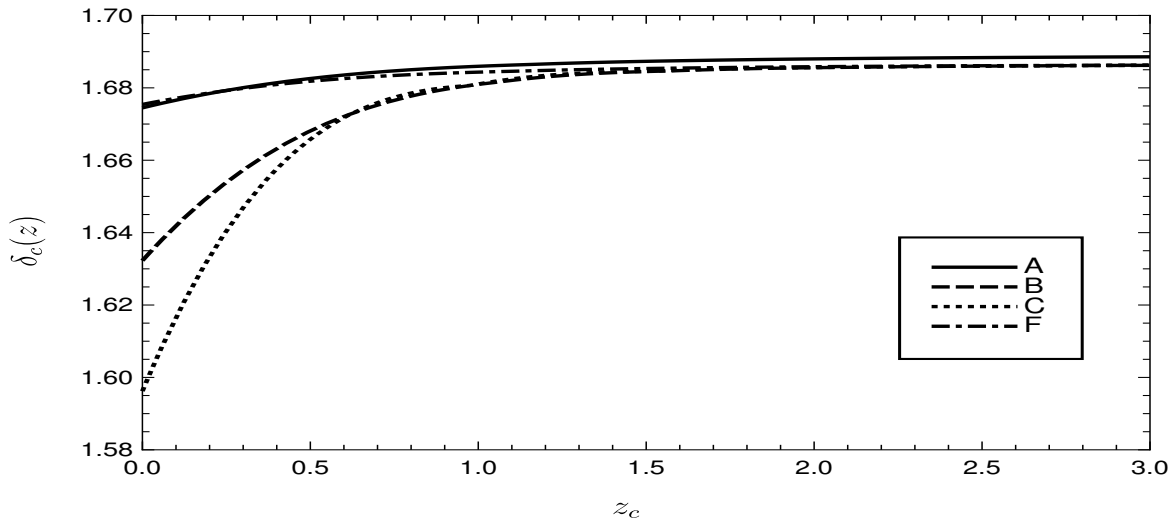}
\caption{\label{fig:3}
Plots of linear density contrast at collapse $\de_c$ as a function of collapsing redshift $z_c$. 
The lines A, B, C and F correspond to models A, B, C and F in Table \ref{modelparam} respectively.
}
\end{figure}

\begin{figure}
\includegraphics[height=0.4\textwidth, width=0.9\textwidth,angle=0]{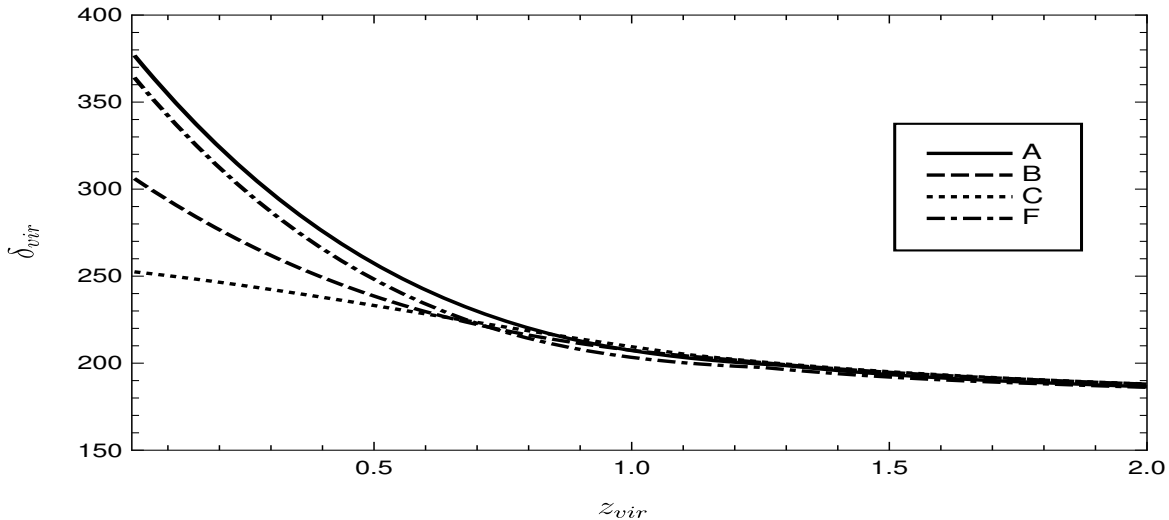}
\caption{\label{fig:deltavirial}
Plots of $\de_{\rm vir}$ as a function of virialized redshift $z_{\rm vir}$.
The lines A, B, C and F correspond to models A, B, C and F in Table \ref{modelparam} respectively.
}
\end{figure}

In spherical collapse model, we are interested in a minimum magnitude of density contrast required for overdensity region to collapse at particular redshift.
This quantity is the extrapolated linear density contrast at collapse which is required in a calculation of mass function of halo. 
To compute this quantity, we numerically solve Eq.~(\ref{depp1}) and search for the initial conditions for $\de_m$ that lead to the collapse, i.e., $\de_m \to \infty$, at a given redshift $z$.
In our calculation, we fix the initial redshift at $z = 10^{5}$ and vary the initial value of $\de_m$ within the range $\de_m \lesssim 10^{-3}$. Hence, we can suppose that initially $\de_m$ obeys linear evolution equation given in Eq.~(\ref{depp1-lin}), and therefore we can set $\de_m' = \de_m$ at initial time.
Then the extrapolated linear density contrast at collapsing redshift $z$, denoted by $\de_c(z)$, is computed by solving Eq.~(\ref{depp1-lin}) from the initial redshift to the collapsing redshift using the initial value of $\de_m$ that lead to the collaps at redshift $z$.
 plots of $\de_c(z)$ are shown in Figure~(\ref{fig:3}). From the plots, we see that increasing the influences from disformal coupling can enhance the decay rate of $\de_c$ at late time. This is a consequence of higher growth rate of density perturbation of dark matter and small energy density of dark energy for disformally coupled model,
i.e., less amount of density perturbations is required for collapsing when growth rate and energy density of matter are large,
which suggests that over dense regions can be efficiently collapsed at late time due to disformal coupling.
The figure also shows that at high redshifts, $\de_c$ for conformally coupled model is larger than that for the other models which is in agreement with \cite{Wintergerst:2010}.
This is a consequent of large energy density of dark energy in conformally coupled model during $1 < z < 3$.
Using the gravitational potential of dark energy derived in \cite{spc1} with an approximation $w_d = -1$,
\begin{equation}
\Phi_d = - \frac{4 \pi G M \rho_d}{5} r^2
\end{equation}
where $M \equiv 4 \pi \rho_m r^3 / 3$ is the total mass of dark matter inside spherical collapsing regions. We compute overdensity of dark matter at virialization $\de_{\rm vir}$ and plot the results in Figure~(\ref{fig:deltavirial}). From the Figure, we see that the overdensity at virialization is suppressed in disformally coupled models compared with pure conformally coupled and uncoupled models. According to our calculation, this is a consequence of low overdensity at turn around in disformally coupled models.

\section{Cluster number counts}
\label{sec5}
\begin{figure}
\includegraphics[height=0.4\textwidth, width=0.9\textwidth,angle=0]{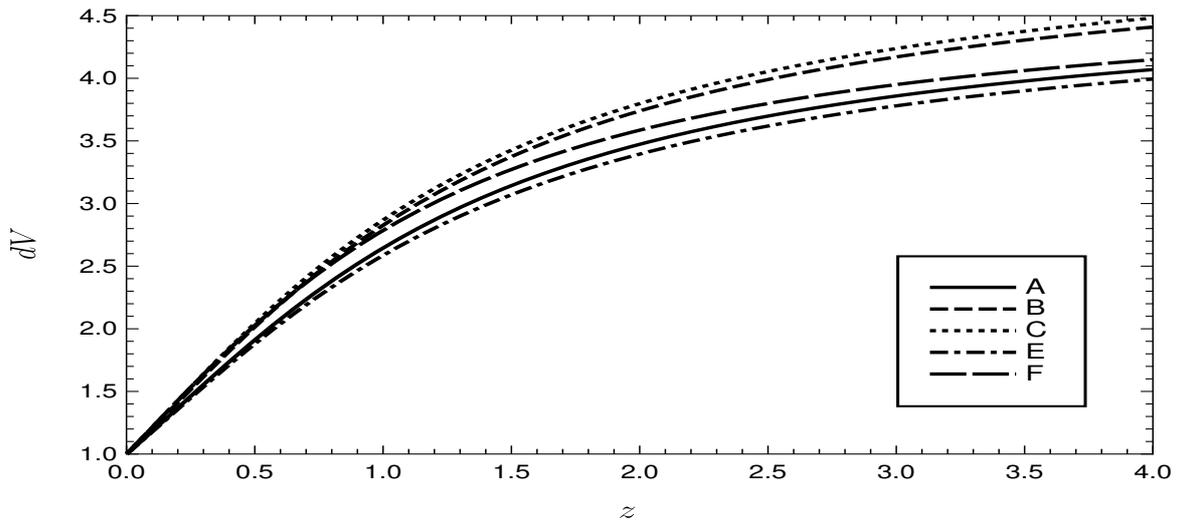}
\caption{\label{fig:4}
Plots of $dV \equiv (dV_e / dZ) / (dV_e / dZ)_{\rm ES}$ as a function of redshift $z$.
The lines A, B, C, E and F correspond to models A, B, C, E and F in Table \ref{modelparam} respectively.
}
\end{figure}

\begin{figure}
\includegraphics[height=0.4\textwidth, width=0.9\textwidth,angle=0]{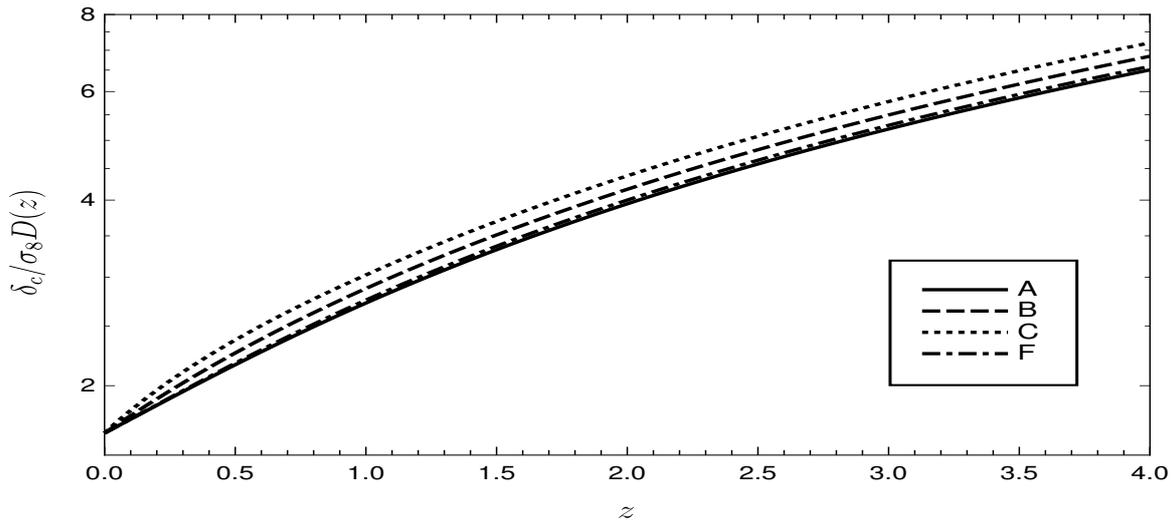}
\caption{\label{fig:5}
Plots of $\delta_c / (\sigma_8 D(z))$ as a function of redshift $z$.
The lines A, B, C and F correspond to models A, B, C and F in Table \ref{modelparam} respectively.
}
\end{figure}

\begin{figure}
\includegraphics[height=0.4\textwidth, width=0.9\textwidth,angle=0]{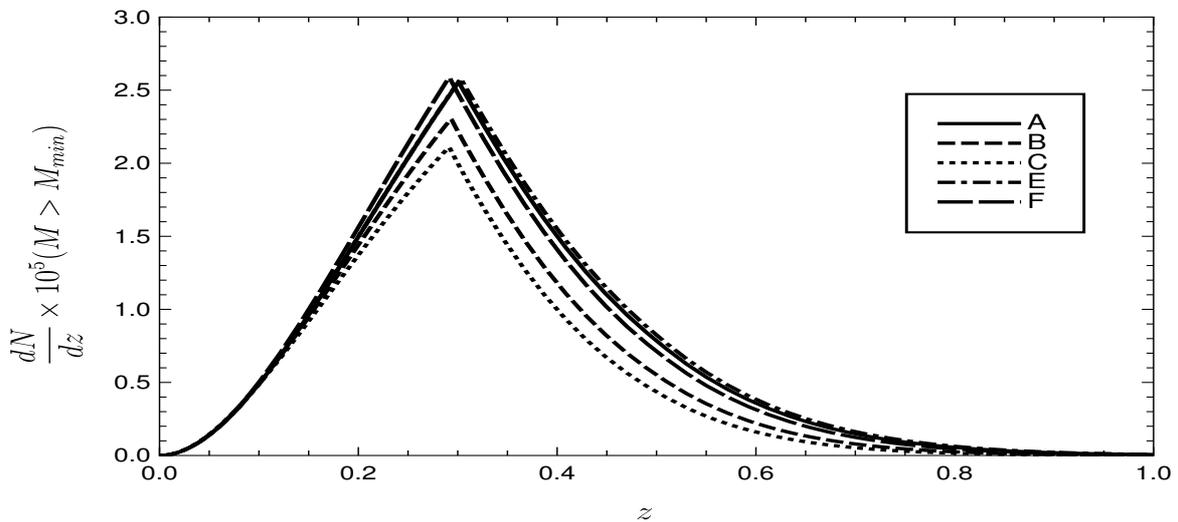}
\caption{\label{fig:6}
Plots of number of cluster per redshift interval $dN/dz$ as a function of redshift $z$. The lines A, B, C and F correspond to models A, B, C and F in Table \ref{modelparam} respectively.
}
\end{figure}

In the Press-Schechter (PS) formalism, the mass function which describes the comoving number density of collapsed objects with mass in the range of $M$ and $M+dM$ is given by
\be
n(M) dM  =
- \sqrt{\frac{2}{\pi}} \tilde{\rho}_m
\(\frac{\delta_c}{\sigma}\) \frac{d \ln \sigma}{d \ln M}
\exp{\(-\frac{\delta_c^2}{2 \sigma^2}\)}  \frac{dM}{M^2}\,,
\label{mfps}
\ee
where $\tilde\rho_m$ is defined in figure (\ref{fig:1}),
$\de_c$ is the extrapolated linear density contrast at collapse computed in the previous section and
$\sigma$ is the variance in spheres of radius $R$ which can be approximately computed from \cite{VianaPS}
\be
\sigma(R,z)=\sigma_8\(\frac{R}{8 h^{-1} {\rm Mpc}}\)^{-\gamma(R)} D(z) \,.
\ee
Here, $D(z) \equiv \delta_m(z)/\delta(0)$ is the growth factor, $\delta(0)$ is the linear density contrast of matter perturbation at present and 
\be
\gamma(R)= (0.3 \Omega_m h + 0.2)
\[2.92+\log_{10}\(\frac{R}{8 h^{-1} {\rm Mpc}}\)\]\,.
\ee
For a better fit with N-body simulation for $\Lambda$CDM, an improved form of mass function is proposed using the assumption of ellipsoidal collapse of halo rather than the assumption of spherical halo collapse in PS formalism. This mass function is the Sheth-Tormen (ST) mass function \cite{SH99.1},
\be
n(M) dM = 
- 0.2709 \sqrt{\frac{2}{\pi}} \tilde{\rho}_m \frac{d \ln \sigma}{d \ln M}
\left(1 + 1.1096 \(\frac{\delta_c}{\sigma}\)^{0.6}\right)
 \exp\left(-\frac{0.707}{2}\(\frac{\delta_c}{\sigma}\)^2\right) \frac{dM}{M^2}\,.
\label{mfst}
\ee
The number of clusters per redshift interval $dz$ with mass larger than threshold mass $M \geq M_{\rm min}$ 
can be computed from the comoving number density of collapsed objects given in Eq.~(\ref{mfps}) or 
(\ref{mfst}) by
\be
\frac{dN}{dz} = f_{\rm sky} \frac{dV_e}{dz} 
\int_{M_{\rm min}}^\infty n(M) dM\,,
\label{dNdz}
\ee
where $f_{\rm sky}$ is the observed sky fraction,
$dV_e/dz \equiv 4\pi r(z)^2/(H_0 E(z))$ is the comoving volume element per unit redshift and
$r(z)$ is the comoving distance. In order to understand influences of the disformal coupling on the cluster number counts, we first study how the disformal coupling affects $dV_e / dZ$ and $\delta_c / (\sigma D(z))$. In Figure~(\ref{fig:4}), we plot the evolution of $(dV_e / dZ) / (dV_e / dZ)_{\rm ES}$ where $(dV_e / dZ)_{\rm ES}$ is the comoving volume element per unit redshift for Einstein-DeSitter model. From the plot, we see that the disformal coupling raises $dV_e/dz$ at all redshift compared with the pure conformal coupling due to large magnitude of the coupling term $\wt{Q_0}$ at late time.
To study effects of disformal coupling on $\delta_c / (\sigma D(z))$, we plot $\delta_c / (\sigma_8 D(z))$ as a function of redshift in Figure~(\ref{fig:5}). For convenience, $\sigma_8$ for each model with different set of parameters $\lm_1, \lm_2$ and $\lm_3$ is set such that the ratio $\delta_c / (\sigma_8 D(z))$ equals to that for $\Lambda$CDM at $z=0$, and $\sigma_8 = 0.83$ for $\Lambda$CDM \cite{CMB-1}. For such setting, the value of $\sigma_8$ for all models, except model D, lies within the 2-$\sigma$ bound from PLANCK 2015 results \cite{CMB-1}. The plots in the Figure~(\ref{fig:5}) show that the ratio $\delta_c / (\sigma_8 D(z))$ for disformally coupled model is larger than that for pure conformally coupled and uncoupled models, which is a consequence of high growth rate of linear density perturbation and low $\de_c$ at late time for disformally coupled models, and $\delta_c / (\sigma_8 D(z))$ equals to that for $\Lambda$CDM at $z=0$.
We now plot the cluster number counts with $M \geq M_{\rm min}(z)$.
In our study, the predicted cluster number counts from PS and ST mass functions present the same features of conformal and disformal couplings on cluster number counts,
so that we plot only the cluster number counts from ST mass function.
In order to make a connection with the results from galaxy surveys, we use the method presented in \cite{Basilakos, massca, Devi:14} to compute $M_{\rm min}(z)$ from limiting flux of the survey.
According to  eROSITA surveys
\cite{erosita1, erosita2}, we set the limiting flux $F_{\rm lim} =3.3\times 10^{-14} {\rm erg s^{-1} cm^{-2}}$
and use a sky coverage $f_{\rm sky} \simeq 0.485$ to plot the expected redshift distribution of clusters in Figure~(\ref{fig:6}).
From the plots, we see that the number of cluster per redshift can be strongly suppressed in disformally coupled models compared with uncoupled model.
Mainly, this is a consequence of large $\de_c / (\sigma D(z))$ in the disformally coupled models.

\begin{figure}
\includegraphics[height=0.4\textwidth, width=0.9\textwidth,angle=0]{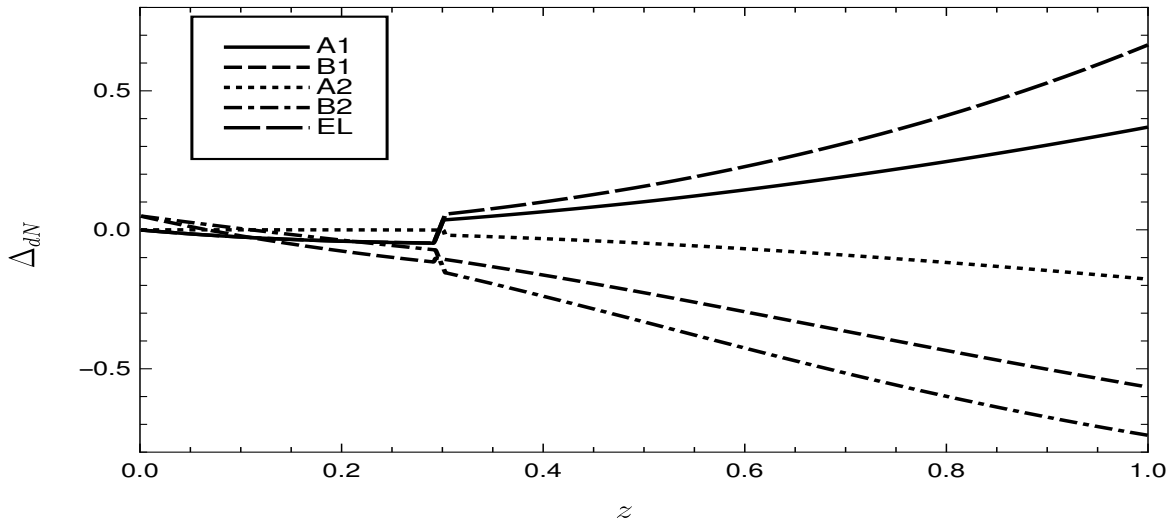}
\caption{\label{fig:7}
Plots of different ratio $\Delta_{dN} \equiv (dN/dz)/ (dN/dz)_f - 1$ from $dN/dz$ presented in the Figure~(\ref{fig:6}). Here, lines A1 and A2 represent the different ratio of line A in the Figure~(\ref{fig:6}) with $\Lambda$CDM and uncoupling model respectively. Lines B1 and B2 represent the different ratio of line B in the Figure~(\ref{fig:6}) with $\Lambda$CDM and uncoupling model respectively. Line E corresponds to the different ratio of uncoupled model with $\Lambda$CDM model.}
\end{figure}

In Figure~(\ref{fig:7}), we plot the different ratio $\Delta_{dN} \equiv (dN/dz)/ (dN/dz)_f - 1$ for disformally and conformally coupled models, where $(dN/dz)_f$ is $dN/dz$ for either $\Lambda$CDM or uncoupled model. From the plots, we see that at $z > 0.3$ the number of cluster per redshift for uncoupled dark energy model is larger than that for $\Lambda$CDM model.
From line A2, we see that the number of cluster is suppressed for all range of redshift by pure conformal coupling between dark energy and dark matter.
However, at $z > 0.3$, the number of cluster for $\Lambda$CDM is smaller than that for pure conformally coupled model because the suppression of cluster number due to pure conformal coupling is not strong enough.
In contrast, the disformal coupling can strongly suppress number of cluster at high redshifts such that the number of cluster for disformally coupled model is always less than that for $\Lambda$CDM and uncoupled models at $z > 0.1$.
The number of cluster can be enhanced at $z < 0.1$ in disformally coupled models due to a large $\wt{Q_0}$ at late time. The enhancement of the number of cluster at low redshifts is mainly a result from a large $dV_e / dz$ in disformally coupled models, so that the choice of $\sigma_8$ does not significantly affect this enhancement. 
Moreover, the features of number counts suppression at high redshifts does not significantly depend how the value of $\sigma_8$ is chosen, in the sense that the conformal and disformal couplings can suppress the cluster number counts and the strong suppression can occur in disformally coupled models.

Combining Figure~(\ref{fig:7}) with Figure~(\ref{fig:6}), we find a difference of $\sim 6850$ clusters at $z \simeq 0.3$ between pure conformally coupled and $\Lambda$CDM models. At $z = 1$, a difference of number of cluster between these models is $158$.
The difference of number of cluster between disformally coupled and $\Lambda$CDM models is $26089$ at $z = 0.3$ and $241$ at $z = 1$ respectively. These differences of number of cluster for disformally coupled model are larger than the estimated eROSITA uncertainty, which are $\Delta N \sim 470$ and $\Delta N \sim 14$ at redshifts $0.3$ and $1$ respectively.
These uncertainties are computed from the Poisson error of the $dN / dz$ for $\Lambda$CDM model plotted in figure (\ref{fig:6}).
The difference of number of cluster between disformally coupled and $\Lambda$CDM models at $z \sim 0.3$ is also larger than the estimated eROSITA uncertainty $\Delta N \simeq 500$ clusters presented in \cite{Devi:14}.
Moreover, at redshifts around the peak of $dN/dz$ the differences of number of cluster between uncoupled model and disformally as well as pure conformally coupled models are also larger than uncertainty of eROSITA surveys.
These suggest that the cluster number counts can be used to distinguish cosmological consequences  of disformal and conformal coupling between dark energy and dark matter and put a tight constraint on disformally coupled models.

\section{Conclusion}
\label{sec6}
We investigate influences of disformal coupling between dark energy and dark matter on large scale structure by using the spherical collapse model and the Press-Schechter/Sheth-Torman mass function to estimate cluster number counts. During matter domination, the disformal coupling has no significant effect on the growth rate of density perturbations of dark matter, so that the collapsing properties of an overdense region (radius, virialisation, critical density) is not altered by this type of coupling. 

The growth rate of density perturbations of dark matter can be enhanced at late time due to a large coupling between dark energy and dark matter in disformally coupled models, as a result, overdense regions can collapse more efficiently at late times, which is suggested by low $\de_c$ at low redshifts. Moreover, the overdensity at virialization in the disformally coupled models can be suppressed at low redshifts compared with conformally coupled and uncoupled models.

Based on the Press-Schechter and Sheth-Torman mass functions, we have found that the predicted number of cluster per redshift interval in disformally coupled models is strongly suppressed compared with conformally coupled and uncoupled models at redshift larger than 0.1  due to a large $\de_c/(\sigma_8 D(z))$.
However, the disformal coupling between dark energy and dark matter can enhance number of cluster at redshift lower than 0.05 due to a large comoving volume element per redshift.
Using the specifications of eROSITA survey, we find that it is possible to discriminate signatures of disformal and conformal coupling between dark energy and dark matter on cluster number counts,
and put tight constraint on disformally coupled models by cluster number counts.

\acknowledgments
KK is supported by Thailand Research Fund (TRF) through grant RSA5780053. DFM thanks the Research Council of Norway for their support and the NOTUR cluster FRAM. This paper is based upon work from COST action CA15117 (CANTATA), supported by COST (European Cooperation in Science and Technology).

\end{document}